# MontiWeb - Modular Development of Web Information Systems

Michael Dukaczewski    Dirk Reiss
Mark Stein
Institut f. Wirtschaftsinformatik
Abt. Informationsmanagement
Technische Universität Braunschweig
http://www.tu-braunschweig.de/wi2

Bernhard Rumpe
Software Engineering
RWTH Aachen
http://www.se-rwth.de

## ABSTRACT
The development process of web information systems is often tedious, error prone and usually involves redundant steps of work. Therefore, it is rather efficient to employ a model-driven approach for the systematic aspects that comprise such a system. This involves models for the data structure that shall be handled by the system (here: class diagrams), various editable and read-only presentations (views) on combinations and extractions of the underlying data (here: a special view language) and ways to connect these views and define data flow between them (here: activity diagrams).

In this paper, we present the MontiWeb approach to model and generate these aspects in a modular manner by incorporating the MontiCore framework. Therefor we shortly introduce the infrastructure that helps to develop modular systems. This involves the whole development process from defining the modeling languages to final code generation as well as all steps in between. We present the text-based class and activity diagram languages as well as a view language that are used to model our system.

## 1. INTRODUCTION
The development of web information systems is a domain that is rather well understood. Quite a number of web application frameworks offer means to implement such systems using a wide range of approaches in almost every modern programming language (for an overview, we refer to [30]). However, most of these frameworks still demand a vast amount of repetitive and tedious work to implement similar parts of a web application: usually a datastructure needs to be implemented following a well-defined and understood scheme, same applies to the persistence mechanisms - either manually written or by using a framework such as JPA [13]. In web systems most of the datastructure need appropriate presentations to provide CRUD (create, read, update and delete) functionality and page flow needs to be defined for each web application. Depending on the technology employed, the effort needed to implement this varies a lot: frameworks like Apache Struts [2] require the maintenance of lengthy and unreadable XML files to specify the flow between different pages.

In order to develop such a system as efficient as possible and thus to reduce laborious and error prone work of manually writing the verbose code and configuration files of a web application framework, the adoption of a model driven approach [21, 15, 14] is usually a good choice. Abstracting from implementation details, the developer can focus on specifying the essentials of the system. These are in particular (1) means to define the data structure of the application, (2) ways that enable the developer to define views on the data structure and (3) the possibility to connect these views and specify the relevant parts of a complete web application. From the models describing these aspects, one or more code generators can create many necessary parts of a web-based system. Of course the discussed languages do not cover every aspect (e.g. complicated authentication or application specific functionality is not covered), but the generators and their frameworks used provide a large part of the basic functionality.

In this paper, we present the web application modeling framework MontiWeb. One of the main targets of this approach is to come up with running prototypes early and refine those in an agile way until the final system is developed. Therefore, the MontiWeb approach does provide defaults. The discussed generators are in particular connected to target frameworks and components, that e.g. do provide persistence and a standard authentication mechanism that however can be replaced and adapted to specific needs.

Generally DSLs can be designed as graphical or text-based modeling languages. Both have its advantages and disadvantages. As we do not focus on graphical frontends, but on agile usability, we use a textual notation due to the advantages presented in [11] and the fact that both can be transformed into eachother.

The rest of the paper is organized as follows: Section 2 introduces the framework we use to implement the web application modeling languages, Section 3 describes the languages in detail, Section 4 presents related work regarding the modeling of web information systems and Section 5 concludes this paper and gives an outlook of future extensions.

## 2. DEVELOPING DSLS USING THE MONTICORE FRAMEWORK
As already mentioned in Section 1, we use the modeling framework MontiCore [18, 17, 19] as technological basis for MontiWeb. MontiCore is being developed at RWTH Aachen and TU Braunschweig. It allows the convenient specification of textual modeling languages and provides an extensive infrastructure to process these. It is designed for the rapid development of domain specific languages. A modeling language can be defined in an integrated format that combines both abstract and concrete syntax in one specification.

```
─────────────── MontiCore-Grammar ───────────────
1  grammar Classviews {
2
3    external Annotation;
4    interface ViewElement;
5
6    Classviews = Annotation* name:IDENT
7      "{" Attributes? Views* "}";
8
9    Modifier = (Editor:["editor"] |
10     Display:["display"] | Field:["field"]);
11
12   View = Annotation* Modifier name:IDENT?
13     "{" ViewElement+ "}";
14
15   ViewParameter implements ViewElement =
16     Annotation* Modifier? name:IDENT ";";
17   // ...
18 }
```

**Figure 1: Definition of AST (Metamodel) and concrete textual syntax for Classviews**

As shown in Figure 1, a grammar in MontiCore starts with the keyword `grammar` and is identified by a name (here: `Classviews`). Non-terminals are notated on the left hand side of a production (here: `Classviews` (6), `Modifier` (9), `View` (12) and `ViewParameter` (15)) and used on the right hand side. Keywords are enclosed in double-quotes whereas named elements have a name in front of a colon, followed by the type of element afterwards (e.g., `name` as the name and `IDENT` as the type of the predefined terminal (6, 12)). Rules can have a cardinality (e.g. `*` (6, 7) for 0 to unlimited occurence, `+` (13) for 1 to unlimited and `?` (7, 12, 16) for optional occurrence) and alternative rules (e.g. (9, 10), seperated by the pipe character (`|`)) are supported. The keyword `external` marks certain non-terminals as defined outside of the actual grammar (3) and needs to be linked to another non-terminal from a different grammar. The keyword `interface` (4) implies that the following element is a placeholder for arbitrary elements that implement this interface. Here, the non-terminal `ViewElement` can be replaced by the non-terminal `ViewParameter` or further here ommitted non-terminals (indicated by the three dots (17)).

Besides these constructs, MontiCore supports extension mechanisms such as grammar inheritance (see [19] for a more detailed description of this and the abovementioned concepts). From the grammar, several tools for model instance processing, model-to-model transformation, and code generation are generated and used within the MontiWeb tool.

## 3. MONTIWEB - MODELING WEB APPLICATIONS

The difficulties with developing web information systems manually were briefly described in Section 1. These problems mainly occur due to the application of different technologies that are not designed to be used together. For data persistence, a relational database management system is the common case. Modern frameworks like Struts [2] or Tapestry [3] use template engines like Velocity [28], Freemarker [9] or XML for generating the presentation. The controller is commonly written in a modern GPL like Java. Since all these technologies are developed independently but still describe the same elements on different levels, changes often need to be made in all of them. For example, if a new attribute shall be added to the data structure, all three layers are affected and need to be modified. Furthermore, often glue code in formats such as XML configuration files need to be touched as well. Thus, a model driven development approach can help a lot in these cases: convenient infrastructure provided, each of these layers can be defined in its own modeling language and describe the appropriate matter concisely. Therefore, adding one field would mainly concern one model element and reflect into all other layers automatically. Here the order in which the models are specified is not important. Modeling can be an incremental process where the different models are written in parallel and independently of eachother and then the consistency between them can be checked on the model level and be ensured through tested code generation. The three modelling languages with syntax and function in the websystem and interaction are described in the following.

### 3.1 Data Structure

The central aspect of a web information system is the underlying data structure. The language describing it should be flexible enough to express all necessary aspects and yet easy and domain specific enough to raise the level of abstraction above manual implementation.

Three requirements for the data structure description are set: (1) A type system (2) composability and (3) relationship between model elements. By a domain specific data type system special characteristics are assigned to the data. Thus validation of data, transformation rules, storage mechanisms and other data-specific functions are easily possible.

Composability of complex data means that one data structure can be made up from elementary data types as well as complex ones defined elsewhere in the model. The relationships between the data define mapping properties. Since class diagrams offer enough expressiveness for data modeling and are generally well-known, MontiWeb uses a textual representation of a subset of UML/P [25, 24] class diagrams to describe the data structure. In the following we explain how the chosen modeling language met the three requirements for the description of the data structure. An example of such notation is shown in Figure 2. It shows the simplified data structure of a carsharing service that consists of persons and cars. A class diagram begins with the keyword `classdiagram` and is named right after (1). It contains class definitions that are notated straight-forward with the corresponding keyword. The different attributes are defined within the class and consist of a type (e.g. `MWString` (4) which represents a domain specific implementation of a String) and a name (e.g. `name` (4)). MontiWeb distinguishes two types of classes: (a) Base classes - are similar to primitive types of Java. They do not include any attributes and are implemented in the target system according to their own rules. (b) Complex classes - contain attributes of base classes as well as other complex classes. To model relation-ships between two classes, associations can be used.

MontiWeb provides two types of associations. Normal associations (not shown in the example) in the generated web system are treated as link between objects, i.e. for an association between class A and B, an object of class A can be assigned to an object of class B. The second type of association is composition. It is denoted by the keyword `composition` (17-18) and the two associated classnames (`Person` and `Car`). Associations can have named roles (`keeper` and `cars`), cardinalities (`*` in this case, the ommission on the other side implies exactly 1) and directions (here, `->` which implies that a person owns cars that only exists in combination with the person). In compositions, one class is embedded into the other class, whereas the embedded object is created simultaneously with its parent object. The composition represents a part-whole or part-of relationship with a strong life cycle dependency between instances of the containing class and instances of the contained classes. This implies that if the containing class is deleted, every instance that it contains is deleted as well. Using to multiplicity and direction, other properties of the association or composition can be defined. In MontiWeb, static selection lists, such as days of the week, can be defined by enumerations (9), and can also be considered as a type of attributes. The entire data structure is distributed over several class diagrams. A class diagram is an excerpt of the overall system. The source code in Figure 2 shows a part of the car sharing web system.

```
                  Classdiagram
1  classdiagram Carsharing {
2
3    class Person {
4      MWString name;
5      Email email;
6      Number age;
7    }
8
9    enum Brand {AUDI, BMW, VW;}
10
11   class Car {
12     Brand brand;
13     Number numSeats;
14     MWDate constYear;
15   }
16
17   composition Person (keeper)
18       -> (cars) Car [*];
19 }
```

Figure 2: Datastructure of a Carsharing application

## 3.2 View Structure

The presentation layer is responsible for rendering the data and providing the interface between a human user and the web information system. Since the main focus of MontiWeb is the domain of data-intensive web applications, the modeling language used offers means to conveniently specify data entry and presentation rather than extensive structures to detailly describe pretty interfaces. Nevertheless, the generated layout can be altered by the common means of adjusting the templates for code generation and the inclusion of Cascading Style Sheets (CSS) and thus fitted to a certain (corporate) design. From a language to specify views of a web system, we demand the following: (a) different, possibly limited views on the underlying data structure must be specifiable, (b) views are composable, i.e. once defined views can be composed to and reused in other ones, (c) static parts (e.g. text or images) can be included in dynamic views on the data and (d) web specific convenience functionality like validation, filtering, sorting data etc. can be modeled with the provided language. Since the UML does not offer any way to specify such features, we developed a domain specific Classview language which allows the specification of different views on a certain class from a class diagram. Each Classview file includes named views on exactly one class (and thus fulfilling the abovementioned requirement (a)). An example of such for class `Person` is depicted in Figure 3.

```
                    Classviews
1  Person {
2
3    attributes {
4      @Required
5      @Length(min=3, max=30)
6      name;
7      @Required
8      age;
9    }
10
11   display protectedMail {
12     name;
13     @AsImage(alt=false)
14     email;
15     cars;
16   }
17
18   display welcome {
19     text {Welcome to Carsharing Service}
20     include protectedMail;
21     age;
22   }
23
24   @Captcha
25   editor registration {
26     name;
27     email;
28     age;
29     cars;
30   }
31
32   display error {
33     @Warning
34     text {You are not old enough!}
35   }
36 }
```

Figure 3: Example of Classviews

Within MontiWeb, special functionality (such as the ones noted above in (d)) is encoded in a syntax that is borrowed from Java annotations. These begin with an ampersand (`@`) and may have additional attribute-value pairs in parens appended to it (e.g. (5)). For MontiWeb, we already offer a rich selection of predefined domain-specific annotations - some of them shown in the example and explained in the following. The rules within the element `attributes` (3-9) apply to all views within the classview file. Here these imply that the attributes `name` and `age` are obligatory to enter (`@Required` (4, 7)) and `name` may appear 3 to 30 chars (`@Length(min=3, max=30)` (5)). These result in the generation of according AJAX verifictaion mechanisms. Subsequently, the different views are specified. These begin with the type of view (here: `display` (11, 18, 32) for views that simply output the data and `editor` (25) that renders the appropriate input fields for the classes' attributes) and are followed by a name. The view `protectedMail` renders the name, email address and cars data of a person whereas the email address is being transformed to an image (caused by the web-specific annotation `@AsImage` to avoid automatic email address harvesting). The `welcome` view displays some static text (19), does furthermore include the `protectedMail` view and displays a persons age. This functionality satisfies the demands (b) and (c) from above. The `registration` view is an editor view and thus provides input fields for `name`, `email` and `age` of a person and – as `cars` denotes the composition of car objects within a person – means to associate such objects to a person. The annotation `@Captcha` (24) produces a captcha field on this view. Finally, the view `error` (32-35) simply consists of a static text that is rendered in a manner that indicates a warning.

An example of how the `registration` view could be rendered is shown in Figure 4.

**Figure 4: View "editor"**

## 3.3 Control- and Dataflow

Defining only the data structure and different views on it suffices for generating basic web information systems that allow rudimentary data manipulation functionality like entering and saving, showing and updating the data. To create more complex web applications, we need means to model both, control and data flow between the different pages or views respectively. For this purpose, we use a profile of UML activity diagrams [22] in textual notation. An example of an activity diagram is shown in Figure 5. It describes a process of user registration where a user enters his user data and is then directed to either a welcome page (in case his age is greater than 18) or an error page (if the age is smaller than 18).

An activity diagram starts with the keyword `activity` followed by the activities' name (here: `UserRegistration`). Actions (introduced by the keyword `action` (3, 8, 13)) posses

```
_________________ Activity Diagram _________________
1  activity UserRegistration {
2
3    action Registration {
4      out: Person p;
5      view : p = Person.registration();
6    }
7
8    action Welcome {
9      in: Person p;
10     view : Person.welcome(p);
11   }
12
13   action Error {
14     in: Person p;
15     view : Person.registrationError(p);
16   }
17
18   initial -> Registration;
19   Registration.p -> [p.age >= 18] Welcome.p
20                   | [p.age < 18] Error.p;
21   Welcome | Error -> final;
22 }
```

**Figure 5: Example of Activity Diagrams**

a name as well and include different contents: `in` (9, 14) and `out` (4) followed by an attribute type (`Person`) and attribute name (`p`) specify input and output parameters of an action. The keyword `view` (5, 10, 15) indicates the kind of content of an action. The view itself is referenced by its name and either can take an object as argument (10, 15) to initialize the view or return an object which is assigned to an output parameter (5). Transitions within an activity are represented by an arrow symbol (`->` (18, 19, 21)) and may contain several sources and targets. The keywords `initial` and `final` denote start and final nodes of an activity and the pipe character (`|` (20, 21)) depicts alternative flows - with conditions on the right hand side (19, 20) or as alternative routes to the final node (21). Object flow is modeled by appending the parameter name to the action name and for simple control flow, these parameters are left out.

Besides these notation elements, concepts such as parallel flow, hierarchical actions (which themselves are specified by an activity) and roles to which actions can be assigned are supported as well but omitted in this paper for the sake of space. Furthermore, different content can be included in an action. Presently, the inclusion of Java code is supported along the already mentioned view calls.

## 3.4 Aggregation (Interaction) of Component Specific Languages

The described models define three views on a whole system. They are developed and specified independently from each other to maintain clean seperation of the different components. Nevertheless the model parts have some well-defined connection points. Elements that are defined in one model are referenced from another (e.g., views are referenced from an action). The inter-model-relationships are essential for completeness and correctness of the whole system and finally define its behavior.

When developing modeling languages from scratch for parts of a domain, first and foremost only these parts are considered. However, although they will work in isolation, they are often used in combination to model the complete system. Therefore, the notation of a language must provide means to connect to other components.

Interaction between modeling languages can be realized with different mechanisms [26], e.g. by embedding one language into another like SQL is embedded into a GPL like Java. In MontiWeb, the inter-language interaction is realized by using the pipeline pattern [26]. There, the different languages are independent but still implicitly connected, i.e. the implicit relationships are explicitly checked within the generation process. The visibility between the MontiWeb models is depicted in Figure 6. The controller functionality is realized similar to the Application Controller pattern [8]. Here, class diagrams are completely independent from the rest of the model. Neither the data presentation (classviews) nor the flow control (activity diagrams) are of importance for the data definition and thus can not be referenced from there. Classviews depend on the data structure as they define explicit views thereof and contain references (e.g. to class-names, attribute names and types or association names) to it. Classviews do not reference activity diagrams, vice versa activity diagrams reference classviews by name. As the control flow defines the central logic of a web information system, both class diagrams and classviews are referenced from there. To maintain consistency between these models, inter-model checks are performed through, e.g. modular symbol tables. Thus the existence of a referenced view or class can be verified.

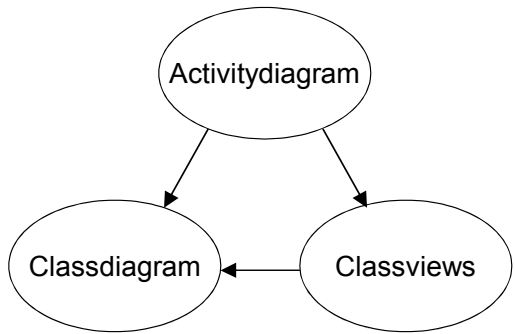

**Figure 6: Models of MontiWeb and their dependencies**

## 4. RELATED WORK

Similar approaches of modeling of web information systems can be classified into (a) modeling using graphical languages, and (b) modeling using textual languages.

One graphical modeling tool in the domain of web information systems is WebML. Unlike MontiWeb, WebML distinguishes two domain segments: (a) data design concerns the specification of data structures and (b) hypertext design is used to describe the structure of web pages [6]. Both of these languages incorporate UML class diagrams. For hypertext design, predefined classes like Entry for the generation of a web form or Data to display a class are used. The navigation structure is depicted by directed associations between classes. Furthermore, WebML supports a XML based textual modeling language which lacks tool support. Therefore, the use of their own graphical modeling tool is favored [5].

UWE (UML-based Web Engineering) [16] follows a similar approach as WebML. It uses class diagrams for data structure specification and, like MontiWeb, uses acivity diagrams to describe the modeling of workflow. UWE's notation is a graphical one as well. Like WebML, the UWE models can be exported in an XML format.

Another tool that uses graphical modeling is AndroMDA [1]. AndroMDA does not have its own editor yet, but uses XMI as input format which is supported by some UML Editors. Like MontiWeb, it uses class diagrams for data structure and activity diagrams for workflow description. AndroMDA does not offer a specific language to describe the view aspect of a web information system, but generates it from extra class diagrams that have to be specified additionally. To get a working application, all parts have to be provided. A generation of standard behavior as MontiWeb does is not supported.

As a textual modeling approach, WebDSL [29] follows a similar approach as MontiWeb. The language there is specified using SDF [12] and Stratego/XT [4] for language transformation. They use a purely domain specific modeling language and is not leaned on UML.

The Taylor project [27] follows an MDA approach to model and develop JEE applications. The models are created using Eclipse Graphical Modeling Framework (GMF) [10] and are stored in XMI format by incorporating EclipseUML [7]. As notation for data structure, Taylor uses class diagrams, business processes are defined by activity diagrams as well. The navigation structure between pages is specified by a state machine language where states depict pages and transitions links from page to page.

Another popular approach for generating web information systems is Ruby on Rails [23]. Although it is not a pure model based approach, a prototype application can be generated using the Ruby on Rails scaffold mechanism. From a simple model in a Rails-specific notation and a HTML-based view template language, CRUD functionality and a very basic controller can be generated. However, unlike MontiWeb, the focus of Rails is the manual programming of all three components, aided by extensive web-specific functionality provided by the language.

The Mod4j (Modeling for Java) [20] project aims at the efficient development of administrative enterprise applications by employing a model driven approach. Like MontiWeb, Mod4j seperates the application into its different aspects and offers a modeling language for each. The Business Domain Model is represented by an UML class diagram. Page flow is modeled using a specific Service Model and the presentation in the application has its own modeling language as well. Mod4j is based on Eclipse technology and uses XText [31] for the development of languages.

## 5. CONCLUSION AND FUTURE WORK

In this paper we described our approach to model and generate web information systems to tackle the insufficiencies that occur when developing such systems manually. Especially the difficulties caused by the combination of normally orthogonal frameworks are approached. Within MontiWeb, we use three languages for the three main segments of a web information system. Two of them come from the UML/P, one language (classviews) is completely new defined. These languages reflect the requirements of each domain component and were adapted to their specific needs.

The currently reached status involves pretty stable languages, as discussed here, and appropriate generation tools. Furthermore a number of presentation forms for various data types (such as Date, String etc. are defined). We currently work on extensions of the provided functionality in various ways. This includes e.g. components for more fine grained security, identification and authentication as well as the possibility to easily integrate predefined (third-party) components that provide application specific functionality. We plan to further extend and complete the already used languages (e.g. include inheritance in class diagrams) and incorporate new ones to model not yet covered aspects of a web information system (such as use case diagrams for requirements modeling). Furthermore, we think of generation of a modular API to access the generated system via SOA-services or add SOA-functionality provided by other servers.